\documentclass[twocolumn,prl,showpacs,amsmath,amssymb]{revtex4}

\usepackage{graphicx}
\usepackage{dcolumn}
\usepackage{bm}

\begin{document}

\title{High precision determination of the $Q^{2}$-evolution of the Bjorken
Sum }

\author{
A.~Deur$^{\njlab}$,
Y.~Prok$^{\nodu, \njlab}$, 
V.~Burkert$^{\njlab}$,
D.~Crabb$^{\nuva}$,
F.-X.~Girod$^{^\njlab}$,
K.~A.~Griffioen$^{\nwm}$, 
N.~Guler$^{\nodu}$\footnote{Present address: 
Los Alamos National Laboratory, Los Alamos,
New Mexico 87544},
S.~E.~Kuhn$^{\nodu}$ and 
N. Kvaltine$^{\nuva}$ } 

\affiliation{
\baselineskip 2 pt
\centerline{{$^{\njlab}$Thomas Jefferson National Accelerator Facility, 
Newport News, VA 23606}}
\centerline{{$^{\nodu}$Old Dominion University,  Norfolk, VA 23529}}
\centerline{{$^{\nuva}$University of Virginia, Charlottesville, VA 22904}}
\centerline{{$^{\nwm}$College of William and Mary, Williamsburg, VA 23187}}
}

\newcommand{\nodu}{2}
\newcommand{\njlab}{1}
\newcommand{\nuva}{3}
\newcommand{\nwm}{4}

\date{\today}

\begin{abstract}
We present a significantly improved determination of the Bjorken Sum
for 0.6$\leq Q^{2}\leq$4.8 GeV$^{2}$ using precise new $g_{1}^{p}$
and $g_{1}^{d}$ data taken with the CLAS detector at Jefferson Lab.
 A higher-twist analysis of the $Q^{2}$-dependence of the Bjorken
Sum yields the twist-4 coefficient $f_{2}^{p-n}=-0.064 \pm0.009\pm_{0.036}^{0.032}$.
This leads to the color polarizabilities $\chi_{E}^{p-n}=-0.032\pm0.024$
and $\chi_{B}^{p-n}=0.032\pm0.013$. The strong force coupling 
is determined to be $\alpha_{s}^{\overline{\mbox{\tiny {MS}}}}(M_{Z}^{2})=0.1123\pm0.0061$,
which has an uncertainty a factor of 1.5 smaller than earlier estimates
using polarized DIS data. This improvement makes the comparison between
$\alpha_{s}$ extracted from polarized DIS and other techniques a
valuable test of QCD. 
\end{abstract}


\maketitle


\section{Introduction}

The Bjorken Sum Rule \cite{Bjorken SR} is a cornerstone in the
study of nucleon spin structure. It has been investigated via polarized
deep inelastic scattering (DIS) at SLAC, CERN, DESY \cite{E142}-\cite{compass}
and Jefferson Lab (JLab) \cite{EG1a/E94010}-\cite{review}. In the limit
of infinite squared four-momentum transfer  $Q^{2}$ the sum rule is \cite{Bjorken SR}:

\begin{equation}
\Gamma_{1}^{p-n}\equiv\Gamma_{1}^{p}-\Gamma_{1}^{n}\equiv\int_{0}^{1}dx\left(g_{1}^{p}(x)-g_{1}^{n}(x)\right)=\frac{g_{A}}{6},\label{eq:bj}\end{equation}
where $g_{1}^{p}$ and $g_{1}^{n}$ are the spin-dependent proton
and neutron structure functions, respectively, $g_{A}$ is the nucleon
flavor-singlet axial charge, and $x$ is the Bjorken scaling variable.
At a finite $Q^{2}$ large enough so that partonic degrees of freedom
are relevant, the Bjorken Sum Rule has been generalized to account
for perturbative QCD (pQCD) radiative corrections (the leading-twist
term) and non-perturbative power corrections (higher-twist terms).
In the $\overline{\mbox{MS}}$ scheme, the sum rule becomes \cite{Kataev}:

\begin{eqnarray}
\label{eq:bj(Q2)} \Gamma_{1}^{p-n}=\frac{g_{A}}{6}\left[1-\frac{\alpha_{s}}{\pi}-3.58\left(\frac{\alpha_{s}}{\pi}\right)^{2}-20.21\left(\frac{\alpha_{s}}{\pi}\right)^{3}+...\right]+ \\ \sum_{i=2,3...}^{\infty}{\frac{\mu_{2i}^{p-n}(Q^{2})}{Q^{2i-2}}}, \nonumber \end{eqnarray}
where the strong coupling  $\alpha_{s}$ has itself the form
of a perturbative series depending on $Q^{2}$, and the $Q^{2}$-dependence
of the higher-twist coefficients $\mu_{2i}^{p-n}(Q^{2})$ is calculable
from pQCD. The logarithmic $Q^{2}$-dependence induced by the pQCD
radiative corrections that dominate for $\alpha_{s}\ll1$ has allowed  QCD 
to be established  as the correct theory of the strong force. In turn, the higher-twist
power corrections $\mu_{2i}/Q^{2i-2}$ characterize QCD in a stronger
coupled regime with typically $\alpha_{s}>0.3$. Here, at lower $Q^{2}$,
partons start to interact strongly and react more and more coherently
to the probing particles. Thus, the higher-twists describe the transition
between the partonic and hadronic degrees of freedom for the strong
force.

The isovector nature of the Bjorken integral makes it a simpler quantity
to understand theoretically than the integrals for the proton or neutron
separately. This is particularly useful for nucleon structure calculations
performed in different $Q^{2}$ ranges that reflect large or small
$\alpha_{s}$. These regimes, with their suitable calculation techniques,
are summarized below.
\begin{itemize}
\item For $Q^{2}$ above a few GeV$^{2}$, the partonic degrees of freedom
are relevant. Here, pQCD can be tested through the leading-twist part
of Eq.~(\ref{eq:bj(Q2)}). The subtraction of $\Gamma_{1}^{n}$ from $\Gamma_{1}^{p}$
removes the nucleon matrix elements $a_{0}$ and $a_{8}$, and provides
a rigorous QCD prediction. The subtraction also cancels the gluon
and quark-singlet contributions to the $Q^{2}$-dependence of the
sum rule. 
\item At intermediate $Q^{2}$ (from a few GeV$^{2}$ down to a few tenths
of GeV$^{2}$), non-perturbative contributions affect the $Q^{2}-$dependence.
Lattice QCD is the leading calculational technique in this regime.
The isovector nature of $\Gamma_{1}^{p-n}$ simplifies lattice calculations
by removing all disconnected diagrams, which are CPU-expensive to
compute \cite{Goeckeler}. 
\item At low $Q^{2}$ (below a few tenths of a GeV$^{2}$), chiral perturbation
theory, which uses effective hadronic, rather than fundamental partonic,
degrees of freedom, is applicable. The suppression of the $\Delta_{1232}$
resonance contribution to $\Gamma_{1}^{p-n}$ facilitates
the chiral perturbation theory calculations, making these predictions
more robust \cite{Burkert Delta}.
\end{itemize}
New data from the JLab CLAS EG1-DVCS experiment, taken on polarized
proton and deuteron targets, have become available \cite{EG1dvcs}.
The kinematics of new data largely overlap the higher $Q^2$ coverage of earlier 
JLab data  \cite{EG1a/E94010}, \cite{the:Bj EG1b}, but with smaller 
statistical errors. On the other hand, the previous JLab polarized data set 
covers lower $Q^2$ and higher $x$. Put 
together with these data, the EG1-DVCS
data allow us to study the Bjorken Sum at higher $Q^{2}$ and with
improved statistical precision. Studies of the earlier data showed
the necessity of precise measurements at moderately large $Q^{2}$,
greater than $\simeq 2$ GeV$^2$,
in order to extract higher-twists, because of the small magnitude
of their total contribution. As Eq. (\ref{eq:bj(Q2)}) suggests, it
may seem to be beneficial to determine higher-twists at lower $Q^{2}$
where the unmeasured low$-x$ contribution to $\Gamma_{1}^{p-n}$
is smaller, the data are more precise, and the higher-twists are enhanced.
However, in the standard perturbative approach, this may not be reliable
due to the following effects:
\begin{itemize}
\item Higher-order twist effects at low
$Q^{2}$ rise quickly and the short $Q^2$-range over which this rise occurs is too small
to disentangle these higher-twists.
\item There is an increasing uncertainty on the twist-2 part because the proximity of
the Landau pole magnifies the uncertainty on $\alpha_{s}$.
\item  While higher orders leading-twist terms are necessary at low $Q^2$, 
the renormalon problem \cite{Ellis Renormalon} jeopardizes the convergence 
of the series and increase the uncertainties due to truncations. 
\end{itemize}
It is possible to avoid part of the
difficulty by developing expressions for the Bjorken Sum Rule with better
convergence properties, as explored in \cite{APT bj expansion}.
We will not pursue this interesting path, and will instead remain consistent
with the previous analyses \cite{EG1a/E94010}, \cite{the:Bj EG1b}
and \cite{d2 E94010}, using the standard expansion, Eq. (\ref{eq:bj(Q2)}),
since the higher $Q^{2}$ kinematics of EG1-DVCS are suited to this
approach.

\section{Analysis}

\subsection{Bjorken Sum\label{sub:Bjorken-Sum}}

\begin{figure}
 \protect\includegraphics[scale=0.44]{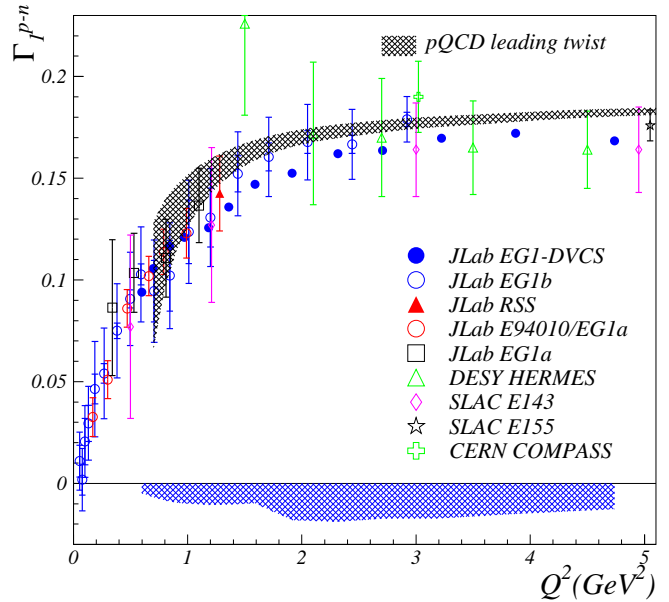}
\caption{\label{fig:bjsr}(Color online.) The Bjorken Sum $\Gamma_{1}^{p-n}$.
The solid blue circles give our results. The blue band is the systematic
uncertainty. Other symbols show the world data. For those, the inner
error bar indicates the statistical uncertainty and the outer error
bar the quadratic sum of the statistic and systematic uncertainties.
The gray band represents the leading-twist NNLO pQCD calculation 
in the $\overline{\mbox{MS}}$ scheme.\protect \\
}

\end{figure}

The extraction of $g_{1}^{p}$ and $g_{1}^{d}$ from the EG1-DVCS
data is described in Ref. \cite{EG1dvcs}. The $Q^{2}$-coverage
and the integration limits are given in Table 1. Since moments must
be integrated over all $x$, a model must supplement the data at low$-x$.
We describe the model  in the next section.
The $Q^{2}$ values for $\Gamma_{1}^{p}$ and $\Gamma_{1}^{d}$ often
differ slightly. When combining them into $\Gamma_{1}^{p-n}$, the
$Q^{2}$ was chosen as the mean between the proton and deuteron $Q^{2}$
values, weighted by the statistical uncertainties on $\Gamma_{1}^{p}$
and $\Gamma_{1}^{d}$. Both $\Gamma_{1}^{p}$ and $\Gamma_{1}^{d}$
were linearly interpolated to the common $Q^{2}$ before being combined
into the Bjorken Sum, $\Gamma_{1}^{p-n}=2\Gamma_{1}^{p}-\Gamma_{1}^{d}/\left(1-1.5\omega_{d}\right)$,
with $\omega_{d}=0.05\pm0.01$ \cite{omega_d} (Here, $\Gamma^{d}$
is calculated as ``per nucleus", not as
``per nucleon"). The result for $\Gamma_{1}^{p-n}$
is plotted in Fig. \ref{fig:bjsr} together with data from the previous
experiments conducted at SLAC \cite{E143}, \cite{E155-E155x},
DESY \cite{HERMES}, JLab \cite{EG1a/E94010}-\cite{the:Bj EG1b},
and CERN \cite{compass}. The elastic contribution ($x$=1) is
not included. Overall, the $Q^2$-behavior of  $\Gamma_{1}^{p-n}$ is smooth within systematic uncertainties. There is good agreement between the world data on $\Gamma_{1}^{p-n}$
and EG1-DVCS, including cases where the neutron moment, $\Gamma_{1}^{n}$,
is obtained from a $^{3}$He target \cite{E154}, \cite{HERMES},
\cite{EG1a/E94010}. We also plot the leading-twist NNLO pQCD
calculation based on Eq. (\ref{eq:bj(Q2)}) (gray band). The width
of the band stems from the uncertainty in the strong coupling 
$\alpha_{s}$. 

\begin{table*}
\begin{tabular}{|>{\centering}p{1.0cm}|>{\centering}p{2cm}|>{\centering}p{1.9cm}|>{\centering}p{1.1cm}|>{\centering}p{1.7cm}|>{\centering}p{1cm}|>{\centering}p{1cm}|>{\centering}p{1cm}|>{\centering}p{1cm}|>{\centering}p{1cm}|>{\centering}p{1.7cm}|}
\hline 
{\small $Q^{2}$ (GeV$^{2}$)} & $x-$range (p) & $x-$range (d) & $\Gamma_{1,meas}^{p-n}$ & {\small $\Gamma_{1,meas+hi.x}^{p-n}$} & $\sigma_{meas}^{syst}$ & $\sigma_{hi.x}^{syst}$ & $\Gamma_{1,tot}^{p-n}$ & $\sigma^{syst}$ & $\sigma^{stat}$ & {\small $\Gamma_{1,meas+hi.x}^{p-n}$}{\small \par}

{\small $/\Gamma_{1,tot}^{p-n}$}\tabularnewline
\hline
\hline 
0.600  & 0.0695-0.072 & 0.070-0.074 & -0.0001 & 0.0612  & 0.0001 & 0.0029 & 0.0940  & 0.0048 & 0.0005 & 0.651\tabularnewline
\hline 
0.698 & 0.0795-0.091 & 0.081-0.094 & 0.0031 & 0.0670 & 0.0002 & 0.0054 & 0.1056  & 0.0068 & 0.0005 & 0.634\tabularnewline
\hline 
0.840  & 0.0970-0.119 & 0.099-0.123 & 0.0079 & 0.0707 & 0.0004  & 0.0079 & 0.1164  & 0.0089  & 0.0006 & 0.607\tabularnewline
\hline 
0.972  & 0.110-0.155 & 0.113-0.168 & 0.0110 & 0.0674 & 0.0008 & 0.0088 & 0.1210  & 0.0099  & 0.0007  & 0.557\tabularnewline
\hline 
1.184  & 0.136-0.210 & 0.139-0.228 & 0.0169 & 0.0628 & 0.0016 & 0.0093 & 0.1257  & 0.0105  & 0.0007  & 0.500\tabularnewline
\hline 
1.361  & 0.151-0.304 & 0.168-0.322 & 0.0414 & 0.0606 & 0.0036  & 0.0082 & 0.1358  & 0.0103 & 0.0009 & 0.446\tabularnewline
\hline 
1.590  & 0.179-0.494 & 0.189-0.494 & 0.0580 & 0.0642 & 0.0083 & 0.0006 & 0.1470  & 0.0098  & 0.0011  & 0.437\tabularnewline
\hline 
1.915  & 0.213-0.804 & 0.233-0.733 & 0.0552 & 0.0542 & 0.0171 & 0.0007 & 0.1524  & 0.0181  & 0.0011  & 0.356\tabularnewline
\hline 
2.316 & 0.263-0.864 & 0.271-0.798 & 0.0523 & 0.0515 & 0.0177 & 0.0001 & 0.1621  & 0.0188 & 0.0008 & 0.317\tabularnewline
\hline 
2.707  & 0.304-0.825 & 0.326-0.769 & 0.0398 & 0.0388 & 0.0157  & 0.0008 & 0.1636  & 0.0173  & 0.0006  & 0.237\tabularnewline
\hline 
3.223  & 0.362-0.901 & 0.385-0.799 & 0.0322 & 0.0311 & 0.0152  & 0.0000 & 0.1697  &  0.0171  & 0.0005  & 0.183\tabularnewline
\hline 
3.871  & 0.438-0.893 & 0.463-0.762 & 0.0227 & 0.0206 & 0.0121 & 0.0002 & 0.1721  & 0.0150 & 0.0004 & 0.120\tabularnewline
\hline 
4.739 & 0.531-0.909 & 0.663-0.738 & 0.0145 & 0.0113 & 0.0081 & 0.0002 & 0.1684 & 0.0126 & 0.0002 & 0.067\tabularnewline
\hline
\end{tabular}

\caption{Kinematic ranges and partial and full Bjorken Sums. 
Columns 2 and 3
give the $x-$ranges over which the proton and deuteron data are measured,
respectively. Column 4 provides  the partial sum  $\Gamma_{1,meas}^{p-n}$ 
from EG1-DVCS. 
 Column 5 gives the measured sum supplemented by a
fit to earlier JLab data in the high-$x$ domain, $\Gamma_{1,meas+hi.x}^{p-n}$.
The experimental systematic uncertainty is denoted by $\sigma_{meas}^{syst}$.
The high$-x$ interpolation is $\sigma_{hi.x}^{syst}$.
Column 8 gives the total $\Gamma_{1,tot}^{p-n}$ sum,  and $\sigma^{syst}$
and $\sigma^{stat}$ are the total (experimental, high$-x$ and low-$x$)
systematics and statistical uncertainties on $\Gamma_{1,tot}^{p-n}$
, respectively. $ $The ratio of the sum without the low-$x$ estimate,
$\Gamma_{1,meas+hi.x}^{p-n}${\small ,} over the total is given by
$\Gamma_{1,meas+hi.x}^{p-n}/\Gamma_{1,tot}^{p-n}$. }

\end{table*}

In order to evaluate the unmeasured parts of $\Gamma_{1}^{p}$ and
$\Gamma_{1}^{n}$ at low-$x$, we need a model for $g_{1}^{p}$ and
$g_{2}^{n}$ covering a wide kinematic range. The model that we use
here is built upon fits to the world data of the asymmetries $A_{1}$
and $A_{2}$, and the unpolarized structure functions $F_{1}$ and
$R$. Those were modeled using a parameterization of the world
data that fits both the DIS and resonance regions with an average
precision of 2 to 3\% \cite{MODEL }. The systematic uncertainty
was calculated by varying either $F_{1}$ or $R$ by the average uncertainty
of the fit (2-3\%) and recalculating all quantities of interest.

For $A_1$ and $A_2$ we used our own phenomenological fit to the world
data, including all DIS results from SLAC, HERA, CERN and Jefferson
Lab and data in the resonance region from MIT Bates \cite{Filoti Ph.D.}
and Jefferson Lab. The asymmetry $A_{2}$ in the DIS region was modeled
using the Wandzura-Wilczek relation \cite{Wandzura-Wilczek}.
For systematic variations, we included a simple functional form for
an additional twist-3 term introduced by E155 \cite{E155-E155x},
and a model constrained by the Soffer Bound \cite{Soffer bound}. 

At very low values of $x$, uncertainties in the model increase rapidly,
so we imposed a lower limit at $x=0.001$. Below this
value, we extrapolate directly the isovector part of the structure
function $g_{1}$ using the Regge parameterization 
$g_{1}^{p-n}(x)=g_{1}^{p-n}(x_{0})(x_{0}/x)^{0.89}$.
We chose the power 0.89 so that the Bjorken Sum at Q$^{2}$ =
5 GeV$^{2}$ from the world data satisfies the Bjorken Sum Rule. Such a 
parameterization agrees within 50\% with the low-$x$ parameterization
determined in Ref. \cite{Low-low-x}. We assumed a 100\% uncertainty
on this contribution. The part below $x=0.001$ contributes up to
about 5\% of the total sum. 

EG1-DVCS does not cover the higher-$x$ values. There, we used a fit to earlier JLab 
data  \cite{EG1a/E94010}, \cite{the:Bj EG1b}.

\begin{figure}
 \protect\includegraphics[scale=0.44]{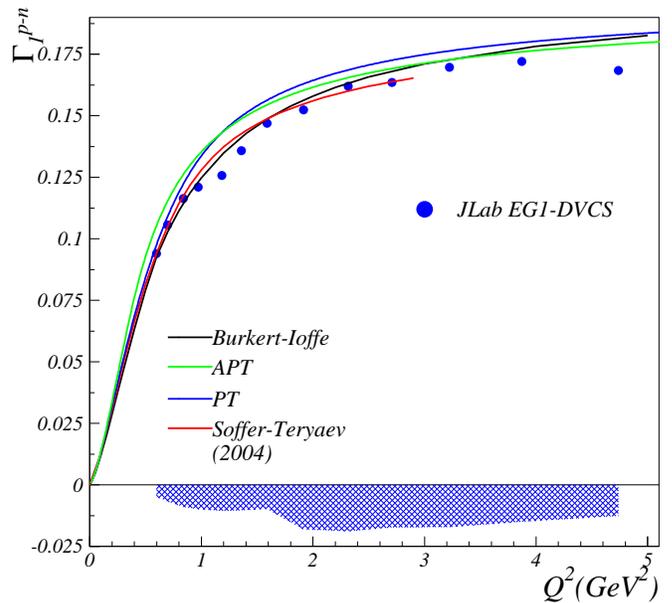}
\caption{\label{fig:bjsr model comp}(Color online.) The Bjorken Sum $\Gamma_{1}^{p-n}$
from EG1-DVCS (solid blue circles) compared with the phenomenological
models described in the main text. \protect \\}

\end{figure}

The new determination of $\Gamma_{1}^{p-n}$ is shown together with
phenomenological models in Fig. \ref{fig:bjsr model comp}. The Burkert-Ioffe
model (black line) is an extrapolation of DIS data based on vector
meson dominance, complemented by a parameterization of the resonance
contribution \cite{AO}. The Soffer-Teryaev model (red line)
uses the smoothness of $g_{1}+g_{2}$ 
with $Q^{2}$ to extrapolate DIS data to lower $Q^{2}$ \cite{soffer1}.
The two other lines are from Ref. \cite{soffer3}. They are updates
of the Soffer-Teryaev model using standard perturbation theory (PT,
blue line) and ghost-free analytical perturbation theory
(APT, green line) which now includes the higher-twist terms $\mu_{4}$
and $\mu_{6}$. The higher-twist values were obtained from fits to
the JLab data \cite{APT bj expansion}. The APT formalism
aims at reducing the influence of the Landau pole divergence at $\Lambda_{QCD}$. 

The precision of the new determination of $\Gamma_{1}^{p-n}$ allows
us for the first time to see that the data lie systematically below
the leading-twist NNLO pQCD prediction shown by the hatched band in
Fig. \ref{fig:bjsr}. Although a large point-to-point correlated contribution
to the systematic uncertainty could still make the data compatible
with the leading-twist calculation, this difference and the steeper
$Q^{2}-$evolution of the data compared to the leading-twist calculation
for $Q^{2}>1.5$ GeV$^{2}$ suggest a negative higher-twist contribution
to $\Gamma_{1}^{p-n}$. These features are quantitatively analyzed
in the next section.

\subsection{Higher-twist analysis}

In this section, we determine quantitatively the higher-twist contributions
to $\Gamma_{1}^{p-n}$. In addition to the EG1-DVCS data, we use all
other world data, including the $Q^{2}=10$ GeV$^{2}$ SMC data \cite{SMC}
not visible in Fig. 1. 

The moment $\Gamma_{1}^{p-n}$ can be expanded in powers of $1/Q^{2}$, see
 Eq. (\ref{eq:bj(Q2)}). The coefficient of the first power correction
is \cite{shuryak}:

\begin{equation}
\mu_{4}^{p-n}=\frac{M^{2}}{9}\left(a_{2}^{p-n}+4d_{2}^{p-n}+4f_{2}^{p-n}\right),\label{eq:HT4}\end{equation}
where $M$ is the nucleon mass. The coefficient $a_{2}^{p-n}$ is
the twist-2 target mass correction expressed as \begin{eqnarray}
a_{2}^{p-n}=\int_{0}^{1}{dx\,(x}^{2}g_{1,LT}^{p-n}),\label{eq:a2}\end{eqnarray}
 in which $g_{1,LT}^{p-n}$ is the leading-twist part of $g_{1}^{p-n}$.
The twist-3 matrix element $d_{2}^{p-n}$ is given by

\begin{eqnarray}
d_{2}^{p-n}=\int_{0}^{1}dx~x^{2}\left(2g_{1}^{p-n}+3g_{2}^{p-n}\right),\label{eq:d2}\end{eqnarray}
and $f_{2}^{p-n}$ is the twist-4 contribution to be extracted. These
coefficients depend logarithmically on $Q^{2}$ but apart for $f_{2}^{p-n}$,
we will neglect this small dependence in our analysis and use their
values at $Q^{2}=1$ GeV$^{2}$. The LO pQCD dependence of $f_{2}^{p-n}$
is accounted for using its anomalous dimension \cite{shuryak}.
The coefficient $a_{2}$ is a kinematical higher-twist \cite{Bluemlein HT}
containing no additional information than is provided by the leading
twist parton distributions. The dynamical higher-twist $d_{2}$ can
be measured directly from polarized lepton scattering off transversely
and longitudinally polarized targets. We are interested here in the
dynamical higher-twist $f_{2}$ which can be obtained only from studying
the $Q^{2}$-evolution of the moment of $g_{1}$.

For a consistent higher-twist analysis, the elastic contribution to
$\Gamma_{1}^{p-n}$ must be added \cite{Ji HT} because it contains
large higher-twist terms, as witnessed by the fast decrease of the
elastic form factors with $Q^{2}$. At $Q^{2}\sim1$ GeV$^{2}$, the
elastic contribution remains sizable and cannot be neglected. To determine
it, we used the elastic form factor fits from Ref. \cite{Arrington elastic FF}
for the proton and Ref. \cite{Mergell FF} for the neutron. The
strong coupling  $\alpha_{s}$ enters in Eq. (\ref{eq:bj(Q2)}).
We computed it in the $\overline{\mbox{MS}}$ scheme to next-to-leading 
order ($\beta_{1}$) in the $\alpha_{s}$'s $\beta-$series. A
fit of polarized parton distributions \cite{BB pdf} was used
to determine $a_{2}^{p-n}$. At $Q^{2}$= 1 GeV$^{2}$, $a_{2}^{p-n}=0.031\pm0.010$.
The proton twist-3 $d_{2}^{p}$ matrix element is obtained from \cite{RSS}.
Data from Refs.  \cite{a1n e99117}, \cite{RSS}, \cite{d2 E94010},~\cite{Solvignon d2},~\cite{Posik d2n} and lattice calculations
\cite{d2 lattice} suggest that for the neutron, $d_{2}^{n}$
is negligible at $Q^{2}>$ 2 GeV$^{2}$. We use $d_{2}^{n}=0.000\pm0.001$
at $Q^{2}=5$ GeV$^{2}$. Evolving $d_{2}^{n-p}$ from $Q^{2}$= 5
GeV$^{2}$ to 1 GeV$^{2}$ using the anomalous dimension calculated
in \cite{shuryak}, we obtain $d_{2}^{p-n}=0.008\pm0.0036$. 

The world data on $\Gamma_{1}^{p-n}$, including those in Table I,
except for the $Q^{2}=$4.7 GeV$^{2}$ point for which the estimated low-$x$
contribution to the integral is large, were fit to extract $f_{2}^{p-n}$
using Eqs. (\ref{eq:bj(Q2)}) and (\ref{eq:HT4}) with $\alpha_{s}$,
$a_{2}^{p-n}$ and $d_{2}^{p-n}$ determined as discussed above. To
account for twist-6 and greater, we add a coefficient $\mu_{6}^{*p-n}/Q^{4}$
to the fit. The asterisk reminds us that this coefficient includes
not only the true $\mu_{6}^{p-n}/Q^{4}$ correction, but also compensations
for higher order terms $\mu_{N}^{p-n}$ with $N>6$. That is, 
$\mu_{6}^{*}$=$\mu_{6}+\Sigma_{i=2,4,..}\mu_{i+6}/Q^{i}$.
The equation shows explicitly that $\mu_{6}^{*}$ depends on $Q^{2}$
(beside its logarithmic dependence that we neglect). Approximating 
$\mu_{6}^{*}$ to be $Q^{2}$-independent is
justified if the power series converges, and this should affect $f_{2}$
minimally but may lead to a $\mu_{6}^{*}$ significantly different
from the actual $\mu_{6}$. We have two completely free parameters,
$f_{2}$ and $\mu_{6}^{*}$, in the fit, plus a third parameter, the
axial charge $g_{a}$, which is bounded by its experimental uncertainty
range ($g_{a}=1.27\pm0.04$).

As published, the world data on $\Gamma_{1}^{p-n}$ are corrected
for the missing low-$x$ contribution using various estimates,
depending on the publication. For the consistency of this analysis,
the low$-x$ estimates of the world data were re-calculated using
the model discussed in the Bjorken Sum section. For all JLab
data sets (Refs. \cite{EG1a/E94010}, \cite{the:Bj EG1b} and
the present data), the point-to-point uncorrelated uncertainties have
been separated from the correlated ones using the \emph{unbiased estimate},
and added in quadrature to the statistical uncertainties. The correlated
systematics were propagated independently into the fit result, as
was the uncertainty arising from $\alpha_{s}$. The uncertainties
stemming from $a_{2}^{p-n}$ and $d_{2}^{p-n}$ are negligible. Table
II gives the best fits for several $Q^{2}$ ranges, since there is
no prescription as to where in $Q^{2}$ the fit should start. The results
are consistent. The first uncertainty listed is the quadratic sum
of the statistical and point-to-point uncorrelated uncertainties. The
second is the point-to-point correlated uncertainty. We do not report
the parameter $g_{A}$ in Table II. Its fit value is always $g_{a}=$1.305,
which corresponds to the upper bound of its variation range. This
is due to the positive elastic contribution that dominates the $Q^{2}-$dependence
of the sum for $Q^{2}\lesssim1$ GeV$^{2}$. For $Q^{2}\lesssim1$
GeV$^{2}$, the $Q^{2}$-dependence of $ $the elastic contribution
is less steep than that of the $1/Q^{4}$ or $1/Q^{6}$ higher-twist terms. 
These too-steep behaviors are compensated in the fits in part by a 
negative $f_{2}^{p-n}$ and in part by an increased leading-twist contribution, 
i.e by a larger $g_{A}$. This compensates for the too-steep $Q^2$-behavior 
of $\mu_6$ or $\mu_8$  compared to the data, since  both the 
leading-twist and the $f_2$ contributions have slopes of opposite signs (their values 
increase with $Q^2$) to that of $\mu_6$ or $\mu_8$ (their values decrease 
with $Q^2$).

\begin{figure}
\includegraphics[scale=0.4]{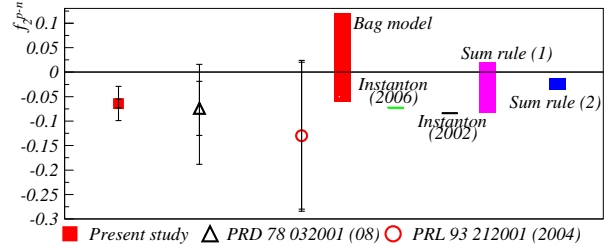}
\caption{\label{fig:f2}Three-parameter fit result for $f_{2}^{p-n}$ from
the present study (square) and Refs. \cite{the:Bj EG1b}
(triangle) and \cite{EG1a/E94010} (circle). The inner
error bar represents the point-to-point uncorrelated uncertainty and
the outer error bar is the quadratic sum of the point-to-point correlated
and uncorrelated uncertainties. Theoretical calculations \cite{stein}-\cite{Weiss 2}
are shown on the right. }
\end{figure}

To assess the convergence of the twist series in Eq. (\ref{eq:bj(Q2)}),
we give in Table III the best fits when an additional $\mu_{8}^{*p-n}/Q^{6}$
coefficient is used (the asterisk has the same meaning as for $\mu_{6}^{*}$).
In these 4-parameter fits, $\mu_{6}$ now gives more properly the
$1/Q^{4}$ power correction. Similar convergence studies were done
in \cite{EG1a/E94010} and \cite{the:Bj EG1b}, and results
for $\mu_{8}^{*p-n}$ were consistent with zero with large uncertainties
ranging from 0.04 to 0.09 depending on the $Q^{2}$ at which the fit
starts. Now, the precision of the data allows us to determine the
magnitude and sign of $\mu_{8}^{*p-n}$. The question of the convergence
of the higher-twist series arises naturally, since Refs. \cite{EG1a/E94010}
and \cite{the:Bj EG1b} indicated that $\mu_{4}^{p-n}$ and $\mu_{6}^{*p-n}$
are of similar magnitudes but opposite signs at $Q^{2} \simeq 1$ GeV$^{2}$.
This suggested a poor convergence of the twist series, at least in
the $Q^{2}$ ranges concerned. With better data, it now appears that
$\mu_{8}^{*p-n}$ and $\mu_{4}^{p-n}$ are of similar size  while
$\mu_{6}^{p-n}$ is small. This indicates
that Eq. (\ref{eq:bj(Q2)}) converges only for $Q^{2}\gtrsim1$ GeV$^{2}$.
The central value of $\mu_{6}^{p-n}$ is significantly smaller than
that of $\mu_{6}^{*p-n}$, once $\mu_{8}^{*p-n}$ is accounted for.
However, $\mu_{6}^{p-n}$ and $\mu_{6}^{*p-n}$ are still compatible
within uncertainties. A systematic study done with the models \cite{AO}
and \cite{soffer1} is described in Ref. \cite{Deur EG1b TN}.
It was performed to better understand the convergence of the twist
series given a truncation at $\mu_{max}^{*}$ (corresponding to $\mu_{6}^{*}$
for the 3 parameter fit and to $\mu_{8}^{*}$ for the 4 parameter
fit) and a choice of $Q_{min}^{2}$, the lowest $Q^{2}$ used in the
fit. The conclusion from the present experimental higher-twist extraction
agrees with the model-based conclusions of Ref. \cite{Deur EG1b TN}:
\begin{itemize}
\item The extraction of $f_{2}^{p-n}$ is stable as $Q_{min}^{2}$ and $\mu_{max}$
are modified in the ranges $0.6\leq Q_{min}^{2}\leq3$ GeV$^{2}$
and $\mu_{6}^{p-n}\leq\mu_{max}^{p-n}\leq\mu_{12}^{p-n}$ 
for the model study, and in the range $0.6\leq Q_{min}^{2}\leq1$
GeV$^{2}$ and with $\mu_{max}^{p-n}=\mu_{6}^{p-n}$ or $\mu_{8}^{p-n}$
for the present experimental study.
\item The coefficient $\mu_{6}^{p-n}$ is small, typically a factor of 6 smaller than 
$f_{2}^{p-n}$ for the model and a factor of 3 smaller for the data, 
although a 3-parameter fit gives a larger $\mu_{6}^{p-n*}$
of similar magnitude to $f_{2}^{p-n}$. Increasing the number of parameters
decreases $\mu_{6}^{p-n}$. This implies the convergence of the series
for $Q^{2}\gtrsim1$ GeV$^{2}$.
\item At $Q^{2} \simeq 1$ GeV$^{2}$, there is an approximate cancellation of
the higher-twist terms (independent of $Q_{min}^{2}$).
\end{itemize}
The overall uncertainty on $f_{2}^{p-n}$ is dominated by the unmeasured 
low-$x$ region. The uncertainty
from $\alpha_{s}$ becomes important only for fits starting at the lowest
$Q^{2}_{min}$ (0.66 GeV$^{2}$) since the effect of the Landau pole becomes
important as $Q$ gets close to $\Lambda_{QCD}$. The JLab data were all
taken with beam energies of up to about 6 GeV. The upcoming 12 GeV
program at Jefferson Lab will significantly reduce this dominant uncertainty
since the measured fraction of $\Gamma_{1}^{p-n}$ above $Q^{2}=2.5$
GeV$^{2}$ will at least double compared to the present measurement
\cite{CLAS 12 GeV}. 
\begin{table*}
\begin{tabular}{|c|c|c|c|}
\hline 
$Q^{2}$ range & $f_{2}^{p-n}$  & $\mu_{6}^{*p-n}$ (GeV$^{4}$) & $\chi^{2}$/d.o.f\tabularnewline
\hline
\hline 
0.66-10.0 GeV$^{2}$ & -0.093$\pm0.006\pm_{0.037}^{0.026}$ & 0.087$\pm0.002\pm_{0.022}^{0.033}$ & 1.03\tabularnewline
\hline 
0.84-10.0 GeV$^{2}$ & -0.064$\pm0.009\pm_{0.036}^{0.032}$ & 0.070$\pm0.004\pm_{0.018}^{0.023}$ & 0.71\tabularnewline
\hline 
1.00-10.0 GeV$^{2}$ & -0.057$\pm0.010\pm_{0.043}^{0.039}$ & 0.065$\pm0.005\pm_{0.019}^{0.021}$ & 0.72\tabularnewline
\hline
\end{tabular}
\caption{Values of $f_{2}^{p-n}$ and $\mu_{6}^{*p-n}$ at $Q^{2}$=1 GeV$^{2}$
from the 3-parameter fit (the parameter $g_{a}$ is not reported in
this table, see main text). The two uncertainties given for $f_{2}^{p-n}$
and $\mu_{6}^{*p-n}$ are the point-to-point uncorrelated (first
number) and point-to-point correlated uncertainties (second numbers).
The last column gives the $\chi^{2}$ per degree of freedom of the
fit, with only the point-to-point uncorrelated uncertainties accounted
for.}

\end{table*}
\begin{table*}

\begin{tabular}{|c|c|c|c|c|}
\hline 
$Q^{2}$ range & $f_{2}^{p-n}$ & $\mu_{6}^{p-n}$ (GeV$^{4}$) & $\mu_{8}^{*p-n}$ (GeV$^{6}$) & $\chi^{2}$/d.o.f\tabularnewline
\hline
\hline 
0.66-10.0 GeV$^{2}$ & -0.044$\pm0.010\pm_{0.054}^{0.055}$ & 0.012$\pm0.010\pm_{0.034}^{0.024}$ & 0.032$\pm0.006\pm_{0.017}^{0.023}$ & 0.63\tabularnewline
\hline 
0.84-10.0 GeV$^{2}$ & -0.035$\pm0.015\pm_{0.041}^{0.037}$ & -0.005$\pm0.020\pm_{0.009}^{0.008}$ & 0.044$\pm0.014\pm_{0.010}^{0.019}$ & 0.66\tabularnewline
\hline 
1.00-10.0 GeV$^{2}$ & -0.020$\pm0.032\pm_{0.031}^{0.025}$ & -0.037$\pm0.032\pm_{0.019}^{0.019}$ & 0.073$\pm0.022\pm_{0.013}^{0.018}$ & 0.67\tabularnewline
\hline
\end{tabular}
\caption{Same as Table II but for the 4-parameter fit. }
\end{table*}
The twist-4 coefficient $f_{2}^{p-n}$ obtained from the 3-parameter
fit over the 0.84-10 GeV$^{2}$ $Q^{2}$ range is plotted in Fig.
\ref{fig:f2} along with the results of Refs. \cite{the:Bj EG1b}
and \cite{EG1a/E94010} obtained using the same fit range, and
theoretical predictions \cite{stein}-\cite{Weiss 2}. %
The magnitude and sign of $f_{2}^{p-n}$ agree with the analysis performed
on $g_{1}(x)$ in Ref. \cite{LSS-HT}, which found that twist-4
corrections to $g_{1}(x)$ are sizeable but change sign at $x\sim0.4$
for the proton, leading to a small integrated value. Our result expressed
as $\mu_{4}^{p-n}/M^{2}=-0.021\pm0.016$ (3-parameter fit with the
0.84-10 GeV$^{2}$ $Q^{2}$ range) also agrees with the several extractions
done in Ref. \cite{APT bj expansion}, which are typically around
$\mu_{4}^{p-n}/M^{2}\sim-0.05$ with a spread of 0.02. Finally, our $\mu_{4}^{p-n}/M^{2}$
is also in agreement with the higher-twists coefficients obtained in \cite{BB2010}, which
after integrating them over $x$ yield $\mu_{4}^{*p-n}/M^{2}=-0.058\pm0.045$.

\subsection{Color electric and magnetic polarizabilities}

The twist-3 and 4 terms of the $\mu_{4}$ coefficient, Eq. (\ref{eq:HT4}),
yield the color electric and magnetic polarizabilities \cite{stein},
\cite{Ji col. pol.}, $\chi_{E}=\frac{2}{3}\left(2d_{2}+f_{2}\right)$
and $\chi_{B}=\frac{1}{3}\left(4d_{2}-f_{2}\right)$ respectively.
Using the value of $f_{2}^{p-n}$ from the 3-parameter fit starting
at $Q_{min}^{2}=0.84$ GeV$^{2}$ and $d_{2}^{p-n}=0.0080\pm0.0036$,
we obtain $\chi_{E}^{p-n}=-0.032\pm0.024$ and $\chi_{B}^{p-n}=0.032\pm0.013$.
The point-to-point correlated and uncorrelated uncertainties on $f_{2}^{p-n}$
were symmetrized and added in quadrature. The polarizabilities are
compatible with those reported in Ref. \cite{the:Bj EG1b} with
a factor of 2 improvement on the uncertainties.

\subsection{The strong coupling $\alpha_{s}$}

The strong force coupling at the Z$^{0}$ pole, $\alpha_{s}(M_{z}^{2})$,
can be extracted from the Bjorken Sum data by solving Eq.~(\ref{eq:bj(Q2)})
for $\alpha_{s}$, and then evolving $\alpha_{s}$ to the Z$^{0}$
pole. However, the relative uncertainty for this method is large,
typically 30\%, and dominated by the model determination of the unmeasured
low-$x$ region. Rather than using an absolute measurement, we can
obtain $\alpha_{s}(M_{z}^{2})$ more precisely by fitting the $Q^{2}$-dependence
of $\Gamma_{1}^{p-n}$ \cite{Altarelli alpha_s}. In our case,
where we include relatively low $Q^{2}$ data points, we must account
for $\mu_{4}^{p-n}$. We can neglect the higher orders since $\mu_{6}^{p-n}$
is small and $\mu_{8}^{*p-n}$ is suppressed as $1/Q^{4}$ compared
to $\mu_{4}^{p-n}$. Since $f_{2}^{p-n}$ was obtained assuming the
validity of the Bjorken Sum Rule and using a theoretical $\alpha_{s}$,
we must use an independent determination of $f_{2}^{p-n}$ to form
$\mu_{4}^{p-n}$. We choose $f_{2}^{p-n}$ from Ref. \cite{Weiss 2},
for which we assumed a 50\% uncertainty. We used a $\overline{\mbox{MS}}$ 
leading-twist expression
of $\Gamma_{1}^{p-n}$ up to order $\alpha_{s}^{5}$ and estimated
the uncertainty due to the truncation of the leading-twist pQCD series
by taking the difference between the $4^{\mbox{\scriptsize{th}}}$ and 
$5^{\mbox{\scriptsize{th}}}$ orders.
We then evolved the extracted $\alpha_{s}$ to the $Z^0$ mass $M_{Z}$ using the
evolution equation up to order $\beta_{3}$ with $\Lambda_{QCD}^{\overline{\mbox{\tiny {MS}}}}=0.214\pm0.070$
GeV. 

Fitting the values of $\Gamma_{1,tot}^{p-n}$ in Table I, starting
at $Q_{min}^{2}$=2.316 GeV$^{2}$ with $g_{A}$ and $\Lambda_{QCD}$
as fit parameters, we obtain 
$\alpha_{s}^{\overline{\mbox{\tiny {MS}}}}(M_{Z}^{2})=0.1123\pm0.0061$.
The uncertainty is dominated by the point-to-point uncorrelated uncertainty
 $\pm0.0050$. The uncertainties from the truncation
of the $\beta$-series and from $a_{2}^{p-n}$, $d_{2}^{p-n}$ and
$f_{2}^{p-n}$ are comparatively small. The point-to-point correlated
uncertainty is $\pm0.0037$, which is dominated by the  low-$x$
estimate. To assess this point-to-point correlated uncertainty,
we separated $\sigma^{syst}$ in Table I into a constant with
respect to $Q^{2}$, which does not contribute to the uncertainty
on $\alpha_{s}$, and a $Q^{2}$-dependent part. The latter is estimated
by calculating  $\Delta\Gamma=d\left(\Gamma_{1,tot}^{p-n}\right)/dQ^{2}\times(Q^{2}$
bin size$)\times(\Gamma_{1,tot}-\Gamma_{1,meas})/\Gamma_{1,tot}$
for each $Q^{2}$ point. For this expression, the relative amount of the unmeasured low-$x$ contribution,
$(\Gamma_{1,tot}-\Gamma_{1,meas})/\Gamma_{1,tot}$ can be obtained
from the last column of Table I. Each $\Delta\Gamma$ is treated as
an additional uncertainty to $\Gamma_{1}^{p-n}$ and is added in quadrature
to the point-to-point uncorrelated uncertainty. 

The Regge exponent determining the (small) contribution to the 
integral below $x=0.001$ was obtained by assuming the validity of the Bjorken 
Sum Rule at $Q^{2}$=5 GeV$^{2}$. This implies evolving Eq. (\ref{eq:bj}) from 
infinite  $Q^{2}$ to $Q^{2}$=5 GeV$^{2}$. In the process, a value for $\alpha_{s}$
must be assumed. However, this initial assumption on $\alpha_{s}$ does not bias 
our determination
of $\alpha_{s}$. The contribution from $x<0.001$ influences
the absolute value of $\Gamma_{1}^{p-n}$ at the few percent level.  Our  $\alpha_{s}$ 
depends on $x<0.001$ only \emph{via} the $Q^{2}$-dependence,  for 
which we assigned the conservative uncertainty just discussed.

Our value of $\alpha_{s}^{\overline{\mbox{\tiny {MS}}}}(M_{Z}^{2})$ is compatible
with the average world data, $\alpha_{s}^{\overline{\mbox{\tiny {MS}}}}(M_{Z}^{2})=0.1185\pm0.0006$,
and it significantly improves the precision on $\alpha_{s}^{\overline{\mbox{\tiny {MS}}}}(M_{Z}^{2})$
from polarized DIS last reported by the Particle Data Group \cite{alpha_s in PDG}. It is
in excellent agreement with the result reported in Ref. \cite{BB2010}, 
$\alpha_{s}^{\overline{\mbox{\tiny {MS}}}}(M_{Z}^{2})=0.1132\pm^{0.0056}_{0.0095}$, 
extracted from the (non-integrated) $g_1$ world data.  Our
result is less precise than direct measurements at the $Z^0$ pole, but has similar precision
to some of the $\alpha_{s}$ results reported by the Particle Data Group. This demonstrates
the viability of determining $\alpha_{s}$ with polarized DIS data, especially since, as
already discussed for $\Gamma_{1}^{p-n}$, the leading uncertainty will be significantly
reduced when the 12 GeV JLab data will become available \cite{CLAS 12 GeV} and
\emph{ a fortiori} if the future polarized EIC becomes available \cite{EIC}.

\section{Summary}

New JLab CLAS data have allowed us to form the Bjorken Sum $\Gamma_{1}^{p-n}$ 
for $0.60<Q^{2}<4.74$
GeV$^{2}$. The sum is consistent with previous JLab data and exhibits
a characteristically strong $Q^{2}$-behavior in the hadron-parton transition
region. The statistical uncertainty is small compared to the systematic
uncertainty, which is dominated by the contribution from the unmeasured
low-$x$ domain. While the  analyses of former JLab data covered the low and
intermediate $Q^{2}$ regions where hadronic degrees of freedom play
a role, the new data  cover the intermediate and partonic
(high $Q^{2}$) domains. This is particularly suited for extracting
higher-twist coefficients and color polarizabilities. These quantities
were extracted from a global analysis of the world data, including
the new JLab data presented in this paper. The twist-4 coefficient
was confirmed to be relatively large in absolute magnitude: 
$f_{2}^{p-n}=-0.064\pm0.036$ compared to the leading-twist 
coefficient $\Gamma_{1}^{p-n,pQCD}=0.141\pm0.013$,
the twist-2 coefficient $a_{2}^{p-n}=0.031\pm010$, and the twist-3
coefficient $d_{2}^{p-n}=0.008\pm0.003$. The net higher-twist effect
is small around $Q^{2}=1$ GeV$^{2}$ because of a cancellation between
twist-4 and the sum of higher power corrections that are of opposite
sign. Fits with four parameters reveal that the twist-6 contribution
is small and the cancellation comes from twist-8 and/or higher contributions.
This implies the convergence of the twist series above $Q^{2}\simeq1$
GeV$^{2}$. The color electric and magnetic polarizabilities were
extracted with a factor of 2 improvement on the uncertainty compared
to earlier analyses. The two polarizabilities are of similar value
but opposite sign. From the $Q^{2}$-behavior of $\Gamma_{1}^{p-n}$
and a model estimate of $f_{2}^{p-n}$, we extracted $\alpha_{s}^{\overline{\mbox{\tiny {MS}}}}(M_{Z}^{2})=0.1123\pm0.0061$.
The precision is a factor 1.5 better than earlier estimates from
polarized DIS, making $\Gamma_{1}^{p-n}$ a viable observable for
determining $\alpha_{s}$. Its agreement with the other $\alpha_{s}$ determined
from different observables provides a consistency check of QCD. 

~

We thank J. Soffer and R. S. Pasechnik for providing the curves from
\cite{soffer3},  J. Bluemlein and H. Boettcher for pointing out the 
importance of threshold matching in the evolution of $\alpha_s$, and  
P.~Bosted for reading the manuscript and for useful discussions. 
This work is supported by the U.S. Department
of Energy (DOE) and the U.S. National Science Foundation. The Jefferson
Science Associates operate the Thomas Jefferson National Accelerator
Facility for the DOE under contract DE-AC05-84ER40150, with additional
support from DOE grants, DE-FG02-96ER40960 (S. K., N. G., Y. P.) and  
DE-FG02-96ER41003 (K. G.).

\end{document}